\documentclass[12pt,preprint]{aastex}

\def\sgr{SGR~1627$-$41}
\def\chandra{{\it Chandra}}
\def\cxo{{\it CXO}}
\def\axp{AXP~1E1841$-$045}

\begin{document}

\title{Precise Localization of the Soft Gamma Repeater SGR~1627$-$41 and the 
Anomalous X-ray Pulsar AXP~1E1841$-$045 with \chandra} 

\author{Stefanie Wachter}  
\affil{Spitzer Science Center, California Institute of Technology, 
MS 220-6, Pasadena, CA 91125, USA}
\email{wachter@ipac.caltech.edu}

\author{Sandeep K. Patel}
\affil{Universities Space Research Association, NSSTC, SD-50, 320 Sparkman
Drive, Huntsville, AL 35805, USA}
\email{Sandeep.Patel@msfc.nasa.gov}

\author{Chryssa Kouveliotou}
\affil{NASA Marshall Space Flight Center, NSSTC, SD-50, 320 Sparkman Drive, 
Huntsville, AL 35805, USA} 
\email{chryssa.kouveliotou@msfc.nasa.gov}

\author{Patrice Bouchet}
\affil{Cerro Tololo Inter-American Observatory, National Optical Astronomy 
Observatory\altaffilmark{1}, Casilla 603, La Serena, Chile}
\email{}

\author{Feryal \"Ozel}
\affil{Hubble Fellow, University of Arizona, 1118 E. 4th Street, Tucson, AZ 85721, USA}
\email{}

\author{Allyn F. Tennant}
\affil{NASA Marshall Space Flight Center, NSSTC, SD-50, 320 Sparkman Drive, 
Huntsville, AL 35805, USA}
\email{}

\author{Peter M. Woods}
\affil{Universities Space Research Association, NSSTC, SD-50, 320 Sparkman
Drive, Huntsville, AL 35805, USA}
\email{}

\author{Kevin Hurley} 
\affil{University of California, Berkeley, Space
Sciences Laboratory, Berkeley, CA 94720-7450, USA}


\author{Werner Becker}
\affil{Max Planck Institut f\"ur Extraterrestrische Physik, 85740 Garching bei M\"unchen, Germany}
\email{}

\author{Patrick Slane}
\affil{Harvard-Smithsonian Center for Astrophysics, 60 Garden Street, 
Cambridge, MA 02138, USA}

\altaffiltext{1}{Operated by the Association of Universities for
Research in Astronomy, Inc., under cooperative agreement with the
National Science Foundation.}

\begin{abstract}

We present precise localizations of \axp\ and \sgr\ with \chandra. We obtained new infrared 
observations of \sgr\ and reanalyzed archival observations of \axp\ in order to refine their
positions and search for infrared counterparts. A faint source is detected inside the error
circle of \axp. In the case of \sgr, several sources are located within the error radius of 
the X-ray position and we discuss the likelihood of one of them being the counterpart. We 
compare the properties of our candidates to those of other known AXP and SGR
counterparts. We find that the counterpart candidates for \sgr\ and SGR~1806$-$20 would have
to be intrinsically much brighter than AXPs to have detectable counterparts with the observational
limits currently available for these sources. To confirm the reported counterpart of SGR~1806$-$20, 
we obtained new IR observations during the July 2003 burst activation of the
source. No brightening of the suggested counterpart
is detected, implying that the counterpart of SGR~1806$-$20 remains yet to be identified.  
\end{abstract}


\section{Introduction}

Soft Gamma Repeaters (SGRs) are sources of brief intense outbursts of
low-energy $\gamma$-rays.  Three of the four known SGRs were
discovered in 1979 (e.g. \citealt{mgg79,mazets79}). The fourth, SGR~1627$-$41, was
discovered in 1998 \citep{kouv98b}; a possible fifth source,
SGR~1801$-$23, has not yet been confirmed \citep{cline00}. SGRs were
recognized as a class of high-energy transients, distinct from the
``classical'' $\gamma$-ray bursts (GRBs) in 1987, on the basis of
their burst recurring episodes, relatively soft $\gamma$-ray spectra,
and very short burst durations (typically $\sim0.1$ s). All SGRs are
sources of persistent low-energy X-ray emission, with luminosities in
the $10^{33}- 10^{36}$ erg s$^{-1}$ range, although SGR burst peak
luminosities reach super-Eddington values between $10^{42}$ and
$10^{44}$ erg $^{-1}$ (see \citealt{kouv04} for a review of the properties of
SGRs).

The cumulative SGR properties point towards young neutron stars as the
sources of their outburst emission, an idea which arose early with the
detection of $\sim 20$ cycles of 8~s periodic hard X-ray flux
oscillations following the March 5, 1979 event from SGR~0526$-$66
\citep{mazets79}.  The subsequent discovery of periodic pulsations in
the steady X-ray flux of SGRs 1806$-$20 and 1900$+$14, at $P = 7.5$
and 5.16~s, respectively, confirmed the neutron star nature of these
sources (\citealt{kouv98a}; \citealt{guid01}; \citealt{hurl99b}).
Moreover, the measurement of the secular increase of these periods
(\citealt{kouv98a,kouv99}), on a very short time scale $P/\dot P =
1500$ years, argues that the rapid spin-down observed for both SGRs is
caused by angular momentum loss from a highly magnetized ($B\sim
10^{14-15}$ Gauss) neutron star, a `magnetar'. The magnetar model was
developed in the early 1990's by R. Duncan and C. Thompson
(\citealt{dt92}; \citealt{td95, td96}).

There is another rare subset of neutron star sources, the Anomalous
X-Ray Pulsars (AXPs), with properties very similar to SGRs,
except for their bursting activity, which had not until recently been
detected. However, detections of SGR like burst emission from
AXPs (\citealt{kasp03}, \citealt{gav03}) now indicate that these objects
might previously have been observed simply at different phases of an
SGR life cycle. There are 6 AXPs known to date. The sky distribution
of SGRs (three on the Galactic Plane, the fourth in the Large
Magellanic Cloud) and AXPs (5 in the Galactic plane and one in the Small
Magellanic Cloud) also indicates that magnetars may be a young source
population. Further, recent observations of SGR~0526$-$66, 1806$-$20, and
$1900+14$ have revealed a cluster of massive stars several arcseconds
away from each SGR source, pointing to a potential massive star
progenitor for SGRs (\citealt{fuchs99}; \citealt{vrba00};
\citealt{corbel03}; \citealt{klose04}). This embryonic association could be strengthened
with deep IR observations of additional precise source positions.

SGR~1627$-$41 was discovered with BATSE on CGRO on 15 June 1998
\citep{kouv98b}; the source emitted over 100 bursts until early August
1998 \citep{woods99}. No further burst emission has been observed up
to date (March 2004). Shortly after its discovery, 
an IPN/RXTE/BATSE error box was constructed
which crossed the Galactic Plane \citep{smith99}. 
This was later refined to a 1.8 degree by 16 arcsecond
IPN error box \citep{hurl99a}.
A search through the SNR catalog of \citet{wg96}
revealed a single SNR consistent with the IPN error box,
G~337.0$-$0.1. A BeppoSAX Narrow Field Instrument (NFI) observation of
the SNR region resulted in the discovery of the X-ray counterpart to
the source, SAX~J1635.8$-$4736, at $\alpha = 16^{\rm h } 35^{\rm m}
49.8^{\rm s}$ and $\delta = -47^{\circ} 35^{\prime} 44^{\prime\prime}$
(J2000) with an error circle of radius 1$^{\prime}$ (95\% confidence
level) \citep{woods99}.  

\axp\ was detected with the {\it Einstein} HRI as an X-ray point source
at the center of the supernova remnant Kes~73 \citep{kriss85}.
A refined
position of the central source was derived using {\it ROSAT} HRI data 
\citep{helf94}. Subsequent observations of this source with {\it ASCA} 
revealed pulsed X-ray emission with a spin period, and period derivative of 
$P\sim11.8$ s and $\dot P \sim 4.73\times10^{-11}$s$^{-1}$, respectively \citep{vas97}.
The equivalent dipolar magnetic field and pulsar spin--down age corresponding to these 
values are $B\sim 8\times10^{15}$ G and $\sim 3900$ yrs, which together with its
spectral properties, classified the source as an AXP. Kes~73 is a shell-type radio SNR with 
no evidence of a plerionic core; its kinematic distance has been determined
to be between 6 and 7.5 kpc, using H I absorption data with the VLA \citep{san92}. 
Although the SNR had been observed with {\it ROSAT}, {\it ASCA}, and {\it BeppoSAX}
(\citealt{helf94}; \citealt{vas97}; \citealt{gott02}), the AXP was not located with 
sufficient accuracy to allow the determination of a reliable optical or IR counterpart. 

We report here refined source locations for \sgr\ and AXP~1E1841$-$045 with an accuracy of
$0.2^{\prime\prime}$ and $0.3^{\prime\prime}$, respectively,  obtained with the \chandra\ X-Ray Observatory
(\cxo). In Section~\ref{s-position} we describe our
search for an infrared (IR) counterpart of the sources and in Section~\ref{s-comp} we
compare and discuss our results to the existing detections and limits
of other magnetar infrared and optical observations in the literature.

\section{Observations}

\subsection{X-rays}

\subsubsection{\sgr}

We observed \sgr\ with \cxo\ on 2001 September 30 (ObsId 1981) for 48.9 ksec and on
2003 March 24 (ObsId 3877) for 25.7 ksec using the Advanced CCD Imaging Spectrometer
(ACIS). The observations and majority of the data reduction are discussed in
detail in \citet{kouv03}. CIAO v2.3 and CALDB v2.21 were used in all
data analysis and processing tasks. Since the second observation was
taken in Very Faint Mode, we were able to utilize the 5$\times$5 pixel
event island to further reduce the ACIS quiescent background (using
{\sl acis\_process\_events}). To avoid detection of spurious faint
sources, we removed residual charge from cosmic ray events and
corrected for the flaw in the serial readout in CCD ACIS-S4 using {\sl
acis\_detect\_afterglow} and {\sl destreak}. Finally, we removed 
the pixel randomization which was applied during standard processing.

To improve the source position accuracy we removed systematic shifts in the 
coordinate systems of the two event lists by shifting 
their aspect solutions until they were consistent with 
each other, as described below. We first applied the recommended 
aspect offsets\footnote[2]{http://cxc.harvard.edu/cal/ASPECT/} for each 
observation, we then created two images ($0.7-7.0$ keV), and used {\sl wavdetect} 
to identify all X-ray sources within 2.2\arcmin\ of \sgr\ in each image.  
We found 5 sources at the same position (within $0.7\arcsec$) in each image, which we used to 
calculate a mean shift between the two observations of $\delta RA=0.66$\arcsec\ 
and $\delta DEC=-0.027\arcsec$. The shift was then applied to the second observation 
data using {\sl dmtcalc} and {\sl reproject\_events}. Finally, we merged the images with {\sl dmmerge}.
Following this process we were able to derive source positions with ACIS-S with an error radius of $\sim 0.6 $\arcsec\ (90\% uncertainty)$^2$. 

\subsubsection{AXP 1E1841$-$045}

A 29.6 ksec ACIS observation of AXP~1E1841$-$045 was retrieved from the \chandra\ archive (ObsId 729) 
and processed in a similar manner to the observations of \sgr. The search for point sources in 
this data set is hindered by the presence of the X-ray bright supernova remnant Kes 73. After application 
of the recommended aspect offsets, the position of the 
central source was determined via a two-dimensional Gaussian fit.  
The central portion of the data is piled up, but this should 
not adversely affect the determination of the centroid. Since we want to improve the overall astrometric 
accuracy through comparison between X-ray and IR source positions (see Section~\ref{s-position}), we also
searched for additional fainter point sources between $0.7-7.0$ keV excluding a circular region of
2\arcmin\ radius centered on the AXP to avoid the SNR Kes 73 contribution.  This resulted
in 26 sources within a 2\arcmin\--6\arcmin\ annulus, choosing a S/N threshold of 2.6.
 
\subsection{Infrared}

\subsubsection{\sgr}

Infrared $JHK_s$ imaging of the field of \sgr\ was obtained on 2001
March 6 with the CTIO 1.5m telescope and OSIRIS and again on 2001
March 17 with the CTIO 4m telescope and OSIRIS. The number of images
acquired, exposure times and coadds are listed in
Table~\ref{t-log}. The pixel scale and field of view (FOV) are 0.461\arcsec\ pixel$^{-1}$,
266\arcsec $\times$ 266\arcsec\ and 0.161\arcsec\ pixel$^{-1}$, 93\arcsec
$\times$93\arcsec\ for the 1.5m and 4m observations, respectively.
Both sets of data were reduced in the same manner using IRAF. First,
we created a median filtered sky image for each photometric band. This
normalized sky was subsequently scaled to the median sky value of each
individual frame and subtracted. Flatfielding was achieved with dome
flats. Finally, the images in each filter were shifted and added,
resulting in one combined image per filter for both the 1.5 and 4m data.

PSF fitting photometry was performed on all of the combined frames
with DAOPHOT II \citep{stet92}.  Since we did not obtain standard star
observations on the nights of the observations, we derived photometric
solutions for our data utilizing the 2MASS All Sky Data 
Release\footnote[3]{http://www.ipac.caltech.edu/2mass/} \citep{skrut01} 
magnitudes for the stars in our field. This proved
somewhat difficult due to the limited spatial resolution of the 2MASS
data in these crowded fields and the fact that most of the bright
sources with the best photometric measurements in 2MASS are saturated
in our images, particularly those obtained on the 4m telescope. The
systematic error from the photometric transformation is 0.08 mag for the
4m and 0.06 mag for the 1.5m data. Limiting $J, H, K_s$ magnitudes for
source detection are $\sim$18.0, 18.0, 17.0 for the 1.5m data and
$\sim$21.5, 19.5, 20.0 for the 4m data.
    
We derived an astrometric solution for both the 1.5m and 4m frames
through comparison to the corresponding 2MASS images and utilizing the
IRAF task ``ccmap''. A total of 75 matched stars were used for the
1.5m images, with a residual in the fit of 0.06\arcsec\ in each
coordinate. For the 4m images, a total of 43 matched stars (due to the
smaller FOV) resulted in fit residuals of 0.06\arcsec\ in each
coordinate.

\subsubsection{AXP~1E1841$-$045}

IR data of the field of AXP~1E1841$-$045 were retrieved from the ESO archive.
Two sets of observations with NTT/SOFI are listed in the archive, 69.D.0339 and 
63.H-0511. We present here only data from 69.D.0339 (see Table~\ref{t-log}
for details) since the observations 
from 63.H-0511 have already been discussed in detail by \citet{meregh01}. 
SOFI was used in large field mode with a 
pixel scale of 0.290\arcsec\ pixel$^{-1}$ and a FOV of 5\arcmin $\times$ 5\arcmin.
The seeing was $\sim$0.8\arcsec.
The data were reduced in an analogous manner to those of \sgr. An astrometric 
solution was again derived through comparison with 2MASS images, matching 107 stars
with a residual fit of 0.06\arcsec\ in each coordinate. We also calculated 
standardized $K_s$ magnitudes through comparison to 2MASS. Since we only have $K_s$
band data, color terms have been neglected in the transformation. The systematic 
error from the fit is 0.05 mag. Our calibrated magnitudes for the stars listed
in \citet{meregh01} are provided in Table~\ref{t-mcomps}. Unfortunately, \citet{meregh01}
do not provide uncertainties for their magnitudes, so a comparison of the photometric 
results is difficult.  
Assuming comparable photometric errors to those in our data set, most star measurements
agree within the combined uncertainties. A few stars however, deviate substantially. 
We consider a difference of $\> 0.5$ mag between the two measurements to be significant and 
all of the relevant stars are marked with a 
* in Table~\ref{t-mcomps}. 
In order to investigate these discrepancies, we reanalyzed the $K_s$ band data set used by
\citet{meregh01}. We performed PSF fitting photometry and calculated standardized $K_s$ 
magnitudes through comparison to 2MASS, 
neglecting any color terms. The resulting measurements are also listed in Table~\ref{t-mcomps}.
In most cases, these redetermined magnitudes agree well with our measurements based on data set 
69.D.0339 (within the uncertainties).  
Star U is affected by a bad pixel in the Mereghetti data set. The data for star 14, 16, 
17, and 22 confirm our results. Some of the very faint sources did not converge in our PSF 
fitting procedure (V, 21, 23). Note, however, that we determine $K_s = 15.73 \pm 0.13$ for the bright star
immediately to the east of star 21, in good agreement with the magnitude listed by 
\citet{meregh01}. Hence the discrepancy in the photometry for star 21 is simply due to the consideration of different sources.

\section{Astrometry}
\label{s-position}

\subsection{\sgr}

To further improve the \sgr\ source position accuracy, we searched our 
combined image of both \cxo\ observations for X-ray sources using the 
source-finding method described in \citet{swartz03}. We detected a total 
of 76 X-ray sources, accepting as detections a signal to noise ratio of 2.6. 
To determine any systematic offsets between our
X-ray and IR astrometric frames, we searched the 2MASS All Sky Catalog
for positional matches to these 76 X-ray sources within a radius of
2\arcsec\ around each source. We found 34 matches, but only considered
the 24 sources within 6\arcmin\ centered on \sgr\ for our offset
calculation, to avoid positional uncertainties introduced by the
distortion of the off--center X-ray point spread function (PSF). The positions of these 24
X-ray sources and the offsets from their corresponding 2MASS counterparts
are provided in Table~\ref{t-source6}. For completeness, we also list
the positions of all 76 detected X-ray sources in
Table~\ref{t-allsources} along with any 2MASS counterparts, as well as
$JHK_s$ photometric data from 2MASS or our 1.5m and 4m images.
Table~\ref{t-allsources} is provided in its entirety only in
the electronic version of the Journal.
After rejection of the largest outliers in the measured offsets ($> 1
\sigma$), we derive an average offset of 
$\delta RA=$0.364\arcsec ($\pm 0.162$) and $\delta DEC=-0.084$\arcsec ($\pm 0.177$)
for the remaining 19 sources. 
We also used the same matched set of positions to
compute a full geometric image transformation (including shifts, scaling and
rotation) between the X-ray and IR images. The resulting best positions for \sgr\ are 16:35:51.828
$-$47:35:23.35 (average shift) and 16:35:51.844 $-$47:35:23.31 (full
transformation) in J2000 coordinates with a 1$\sigma$ error of 0.2\arcsec\ in each coordinate. 
Both positions lie well within the IPN error box for \sgr\ \citep{hurl99a}
and are also consistent with the positions of
the BeppoSAX and ASCA X-ray sources \citep{woods99,hurl00}.
We are therefore confident that we have identified the X-ray counterpart
of \sgr.

\subsection{AXP~1E1841$-$045}

\begin{figure}[tb]
\epsscale{1.00} 
\plotone{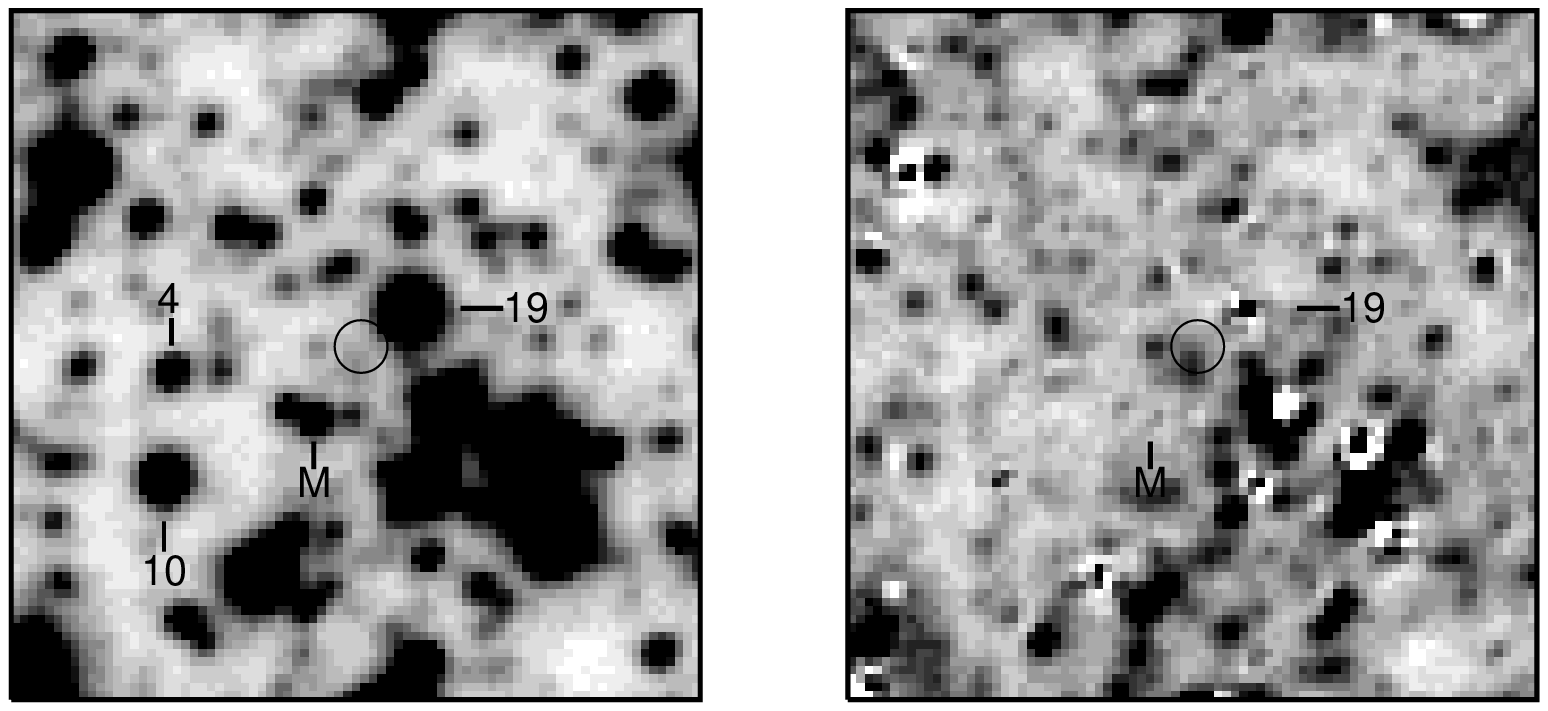} 
\epsscale{1.00} \caption{{\it Left:} The best
\chandra\ localization (3$\sigma$ error radius of 0.9\arcsec) for \axp\ overplotted on our
ESO $K_s$ band image. North is up, east to the
left. Nomenclature of marked stars is according to \citet{meregh01}. {\it Right:} The same field after PSF subtraction of the brighter stars. A faint 
source is visible inside the error circle of \axp. 
\label{f-fcaxp} }
\end{figure}

Our initial best \chandra\ position for \axp\ is RA = 18:41:19.336, DEC = $-$04:56:10.83 (J2000). 
We detected 26 sources in the 2-6\arcmin\ annulus around the central position 
of AXP~1E1841$-$045 above a S/N threshold of 2.6. As for \sgr, we searched for 
matches with 2MASS sources and found 8 sources within 2\arcsec\ of their corresponding 
X-ray positions.  
After excluding any blended sources (based on the higher resolution 
ESO images) we are left with the 5 sources listed in Table~\ref{t-sourceaxp}.
In this case, the number of matched sources is insufficient to derive a full astrometric transformation
between the X-ray and IR frames. Consequently, we only calculated median offsets arriving at  
$\delta$RA=$-$0.10\arcsec\ and $\delta$DEC=0.33\arcsec. 
Including this offset, our final best position of AXP~1E1841$-$045 is  
RA = 18:41:19.343, DEC = $-$04:56:11.16 (J2000) with a 1 $\sigma$ error of 0.3\arcsec\ in each coordinate.  
The left panel of Figure~\ref{f-fcaxp} shows our final \axp\ \cxo\ position with a 3$\sigma$ 0.9\arcsec\
error radius overlaid on the ESO $K_s$ band image (23\arcsec $\times$ 23\arcsec). North is
up, east to the left. A few stars are numbered according to the convention used in 
\citet{meregh01} to facilitate comparison between the two finding charts. 

Star 19, the ``promising
candidate'' identified by \citet{meregh01} appears to be excluded by our refined
X-ray position.
Instead, a very faint
source is visible inside the error circle.  
After PSF subtraction of the brighter stars in the field (right panel of Figure~\ref{f-fcaxp}),
this faint source appears to be a blend of at least two sources. PSF fitting photometry 
proved difficult. Our best fit consists of a brighter southern source ($K_s=17.39 \pm 0.3$) and
a fainter northern source near the detection limit of the observations ($K_s=19.4 \pm 0.5$).
If \axp\ is associated with the SNR Kes 73, the distance to the source is 6--7.5 kpc 
\citep{san92}. \citet{morii03} determined $N_{\rm H}=2.54\times10^{22}$ cm$^{-2}$
from the Chandra data
which translates into $A_V$=14.2 using the relationship by \citet{ps95}. This results in 
$K_0 = 15.77$ and $K_0 = 17.78$ for the brighter and fainter source, respectively. Based 
on these dereddened magnitudes, it is unlikely that the brighter source (which is also 
further away from the center of the error circle) is a viable counterpart candidate as it 
would make it the brightest known counterpart by almost a magnitude (for a detailed 
comparison of the AXP counterpart candidates see Section~\ref{s-comp}). The fainter source, 
however, is promising. Unfortunately, the archival observation we analyzed only contains
$K_s$ band data which prevents us from investigating the color of the candidate. 
An IR color-magnitude diagram for this field has been presented in \citet{meregh01}, 
however, those
data do not go deep enough in $J$ to determine the brightness of our proposed counterpart.      

\section{The Environment of \sgr}

\begin{figure}[tb]
\epsscale{1.00} 
\plotone{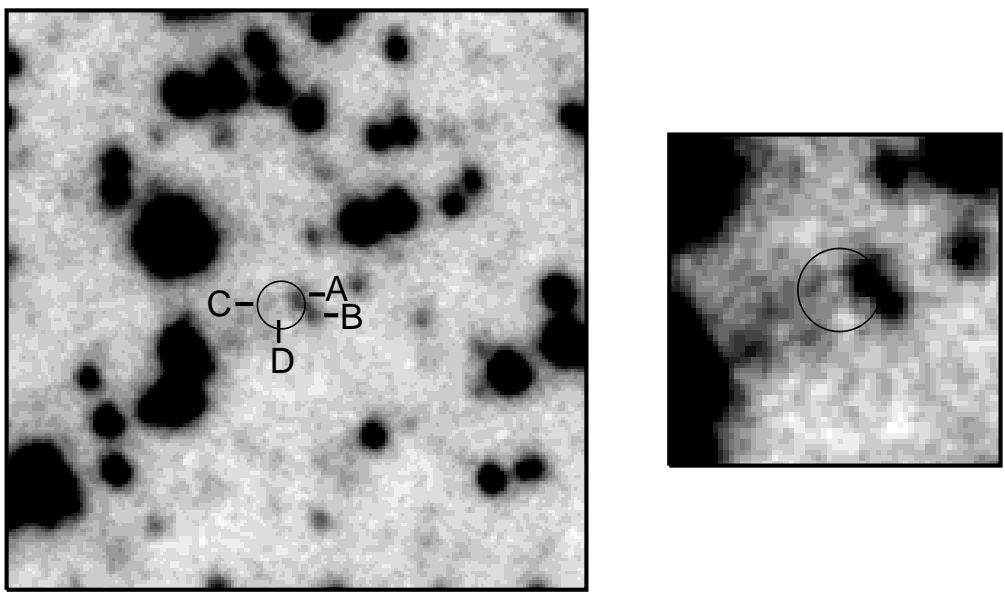} \epsscale{1.00} \caption{The best
\chandra\ localization (3$\sigma$ error radius of 0.6\arcsec) for \sgr\ overplotted on our
deepest CTIO 4m $K_s$ band image ({\it left:} 15\arcsec $\times$ 15\arcsec field of view; {\it right:} close-up of 
the immediate vicinity of the X-ray error circle). North is up, east to the
left.  The sources closest to the X-ray position are labelled.
\label{f-fc} }
\end{figure}

Figure~\ref{f-fc} shows the best \sgr\ \cxo\ position with a 3$\sigma$ 0.6\arcsec\
error radius overlaid on our CTIO K$_s$ band image (15\arcsec $\times$ 15\arcsec). North is
up, east to the left. Despite the excellent seeing conditions during
the observations ($\sim 0.5$\arcsec), it is unclear whether the
sources indicated as A and B are really just two sources or a composition of
multiple fainter sources. Photometric measurements for A and B are
given in Table~\ref{t-phot}. Unfortunately, sources C and D were too
faint to be measured in any of the filters. Source D is closest to the
center of the \cxo\ X-ray error circle, but it is uncertain whether
it is a real source or simply a noise spike.

\begin{figure}[tb]
\epsscale{1.00} 
\plotone{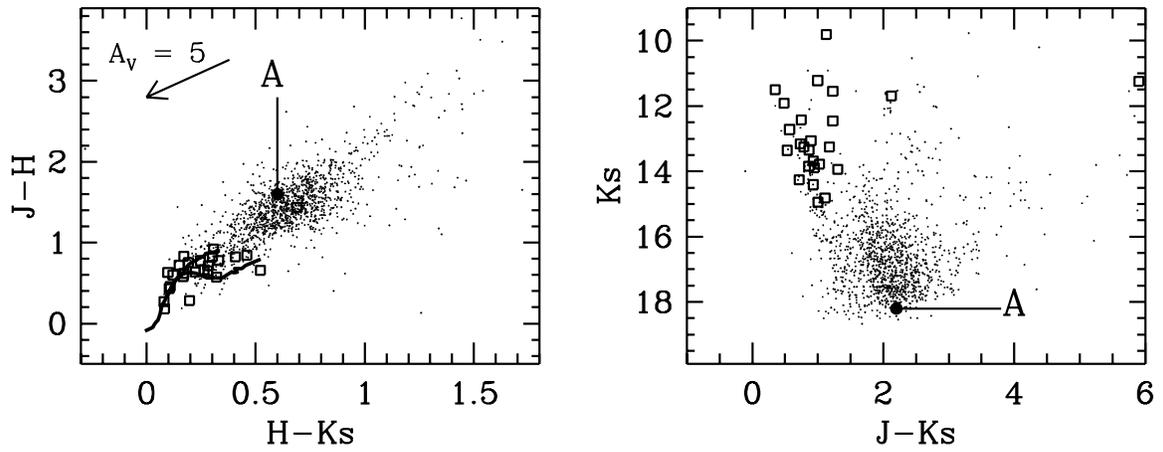} \epsscale{1.00}
\caption{Color--color ({\it left}) and color--magnitude ({\it
right}) diagrams based on the 1296 sources from our \sgr\ CTIO 4m $K_s$ band
image that were detected in all three filters. Open squares indicate
the X-ray sources with 2MASS counterparts, the solid lines in the color--color 
diagram trace the loci of unreddened main sequence (lower fork) and giant stars (upper fork).
\label{f-color} }
\end{figure}

We constructed an IR color--color as well as color--magnitude diagram
from the 1296 sources in the 4m image that were detected in all three
filters (Figure~\ref{f-color}). 
We also indicated the location of the IR
counterparts of {\it all} X-ray sources (not just those within 6\arcmin\ of \sgr) 
with open squares. Of course, not
all of these sources fall in the FOV of our 4m frame,
however, the color--color and color--magnitude diagrams are not
expected to change drastically over the \cxo\ FOV, so that the 
positions of these sources are indicative of
the type of sources we are dealing with. Only those IR counterparts with
accurate photometric measurements, either from 2MASS (errors other
than {\it null}) or our 1.5m and 4m images are shown. The solid lines in the
color--color diagram trace the loci of unreddened main sequence (lower fork) and
giant stars (upper fork). The giant and main sequence colors up to a spectral type
of M6 are taken from \citet{bb88} and transformed to the 2MASS
photometric system with the relations given in \citet{carp01}.  We
also added the 2MASS colors for late type M stars listed by
\citet{giz00}.

The different branches visible in the color--magnitude diagram
distinguish different types of stars. The first branch, roughly up to
$J-K_s = 1.8$, corresponds to nearby main sequence stars, while the
clump of stars around $J-K_s = 2.0-2.5$ represents a superposition of
giant stars with different values of extinction and distance. Most of
the X-ray sources with IR counterparts are likely to be X-ray active
nearby normal stars, since they fall on the unreddened main sequence
and giant star tracks in both the color--color and color--magnitude
diagrams. There is one extremely red source ($J-K_s\sim6$)
corresponding to source 42 of Table~\ref{t-allsources}. It is
undetected in $J$, but bright in both $H$ and $K_s$. The source was in the \cxo\ 
FOV only during the second observation. The very few source counts 
(76 total between $0.5-8.0$ keV) were fitted equally 
well with an absorbed power law and a thermal model, with 
$N_{\rm H}=7.8^{+5.5}_{-3.5}\times10^{22}$ cm$^{-2}$, $\gamma = -4.3^{+2.1}_{-0.7}$, and 
$N_{\rm H}=5.7^{+4.7}_{-1.8}\times10^{22}$ cm$^{-2}$, $kT=1.7\pm0.8$, respectively. 
Most likely this source is an unusually red AGN \citep{wilkes02,cutri01}. 

If we consider the IR magnitudes of the two most secure counterpart 
identifications, AXP~0142+614 \citep{israel03b} and AXP~1048.1-5937 \citep{wang02},  
and place these sources in the IR color--color diagram, they occupy a distinctly
different location than source A and the majority of the field stars. With 
$J-H=1.2, H-K=1.0$ and $J-H=0.9, H-K=1.4$, respectively for the two sources, 
they are much redder in $H-K$ than the bulk of the stars in the frame. Correcting
for the extinction of $A_V\sim 5$ shifts them in the direction of the indicated 
vector to the red edge of the end of the unreddened main sequence. On 
the other hand, reddening the intrinsic colors of AXP~1048.1-5937 with the 
$A_V\sim50$ for \sgr\ would result in $J-H=4.5, H-K=3.9$, shifting its location
off the displayed color--color diagram.
The position of source A in these diagrams shows that the source has normal colors 
consistent with some type of highly reddened
giant star and can thus be rejected as a counterpart candidate for \sgr.

\sgr\ is located at a distance of 11.0$\pm$0.3~kpc in the radio
complex CTB 33, which consists of HII regions and the SNR~$G337.0-0.1$
\citep{sarm97}.  Initially, the source was thought to be associated with the SNR, lying
$\sim $30\arcsec\ outside its rim. However, this association has been
questioned by \citet{gaens01}, who suggested that both the SNR and \sgr\
are simply part of the same star forming complex.  \citet{corb99}
reported on millimeter observations of the line of sight to \sgr, and
concluded that SNR~$G337.0-0.1$ is interacting with a massive giant
molecular cloud (MC-71).  Both \sgr\ and the SNR appear to lie on the
near side of the molecular cloud, since the total extinction up to
MC-71\footnote[4]{Including MC-71 would increase the expected
extinction to $A_V=81.1\pm4.5$ mag} deduced from the millimeter observations is $A_V=39.7\pm2.5$
mag, in good agreement with the $A_V=43\pm3.9$ mag extinction derived
from the N$_H$ absorption of the \sgr\ Beppo-Sax X-ray spectrum
\citep{woods99}. Fits to our most recent \cxo\
observations of the source give $N_{\rm H} = 9.6\pm 1.1 \times 10^{22}$ atoms cm$^{-2}$
\citep{kouv03}.  Using the relation by \citet{ps95} this Hydrogen column density 
translates to $A_V=54\pm 6$, slightly larger than the previous
determination.

To determine whether any of the sources visible inside the 
\cxo\ error circle are viable counterpart candidates for \sgr, we have 
studied the brightness and colors of IR counterparts
identified for other magnetars. We summarize the properties of the
known IR counterparts in the following section.

\section{Counterparts of other SGRs and AXPs}
\label{s-comp}

Currently, four AXPs have convincing IR counterparts while
none are known for SGRs. \citet{eiken01} reported the presence of a
faint IR source inside the \cxo\ error circle for SGR~1806$-$20,
but it remains to be confirmed whether this is the actual counterpart
or a chance superposition in the crowded field of the source. The IR
brightness of several AXPs has been determined on more than one
occasion and appears to indicate significant variability. Particularly
during bursting phases, the IR brightness has been seen to increase
dramatically \citep{kasp03}.

We have compiled the IR photometric measurements, distances, and
extinction estimates for SGRs and AXPs available in the literature in
Table~\ref{t-axpsgrs}. In order to compare our IR sources to the
established AXP counterparts, we corrected their published IR
magnitudes for extinction using the relation between $A_V$ and $A_K$
by \citet{card89}. The resulting intrinsic $K_0$ magnitudes are listed
in the 5th column of Table~\ref{t-axpsgrs}. Note that no effort was
made to correct for the difference in distances, since the distances to
these objects are very uncertain. Among the AXPs, $K_0 = 19-21$
appears to be typical with the exception of counterpart candidate A for 1RXS J1708$-$4009.
\citet{israel03} give two, rather discrepant, $K$ measurements for candidate A. 
The origin of this difference is not discussed.
In any case, candidate A would be by far the intrinsically brightest AXP counterpart. 
\citet{hull04} derived predicted $K_s$ magnitudes based on the assumption that all
AXPs have similar IR to X-ray flux ratios. They argue that candidate B for 1RXS J1708$-$4009
is more likely to be the true counterpart based solely on its brightness, while
\citet{israel03} prefer candidate A due to the similarity of its $H-K$ color to other AXP
counterparts. Our counterpart candidate for \axp\ is also intrinsically brighter than most
of the other AXPs and \citet{hull04} predict substantially fainter IR fluxes for this source. 
Deep, multicolor imaging of both 1RXS J1708$-$4009 and \axp\ are needed to investigate the 
intrinsic range of IR fluxes between AXPs.  
 
Since AXPs and SGRs appear to be manifestations of the same type of
object, we might expect that the counterparts of SGRs would be of comparable brightness
to those of AXPs. Our Table~\ref{t-axpsgrs} shows that the detection limits
of current SGR IR observations would not be sufficient to detect sources of the 
same intrinsic brightness as the known AXP counterparts.  
Only observations of SGR~1900+14 have even approached
the magnitude limits necessary to detect a counterpart with the
intrinsic brightness typical for an AXP. Consequently, the current IR candidates
are unlikely to be the true counterparts for \sgr\ and SGR~1806$-$20, unless
SGRs are intrinsically much brighter than AXPs.

In an effort to answer this question by confirming  
the counterpart candidate for SGR 1806$-$20, we have observed the sources inside the
\cxo\ error circle reported by \citet{kap02a} for SGR~1806$-$20 during recent burst
activity in July 2003. We obtained IR $K_s$ band observations of the field 
on 2003 July 31 with the ESO NTT and SOFI (exposure time and coadd
parameters are listed in Table~\ref{t-log}), only 17 days
after renewed burst activity was detected from the source
\citep{hurley03}.  The data were reduced and analyzed in the
identical manner as our \sgr\ observations. We searched for brightening
of any of the candidate counterpart sources proposed by
\citet{eiken01} which would uniquely identify the SGR counterpart.
The seeing of our image was inferior to that of \citet{eiken01} and we
could not determine the brightness of all of the sources individually.
In particular, we could not resolve sources A and B (nomenclature from
\citealt{eiken01}) and only measured a combined magnitude for these
sources.  The resulting photometry is listed in Table~\ref{t-1806}
together with the photometry given by \citet{eiken01}. Photometry for
A+B was very difficult to obtain due to the vicinity of several very
bright sources.  The combined magnitudes for components A and B
between our and Eikenberry's observations are consistent with each
other within the errors and do not support any significant brightening
of one of the sources. Hence, the counterpart for SGR~1806$-$20 remains 
unidentified. 

\section{Conclusions}

We have determined precise localizations of \axp\ and \sgr\ with \chandra allowing  
for the first IR counterpart search for \sgr\ as well as a refined
search for the counterpart of \axp. We
compare the properties of our candidates to the detections and limits
of other known AXP and SGR counterparts reported in the literature.
Three out of the
four SGR sources are heavily absorbed; optical observations of
the fourth source, SGR~0526$-$66, have not revealed a counterpart
despite the low extinction towards that source \citep{kap01}. We
find that the remaining SGRs would have to be intrinsically much
brighter than AXPs to have detectable counterparts in the observations
currently available for these sources. 

\begin{figure}
\epsscale{0.80}
\plotone{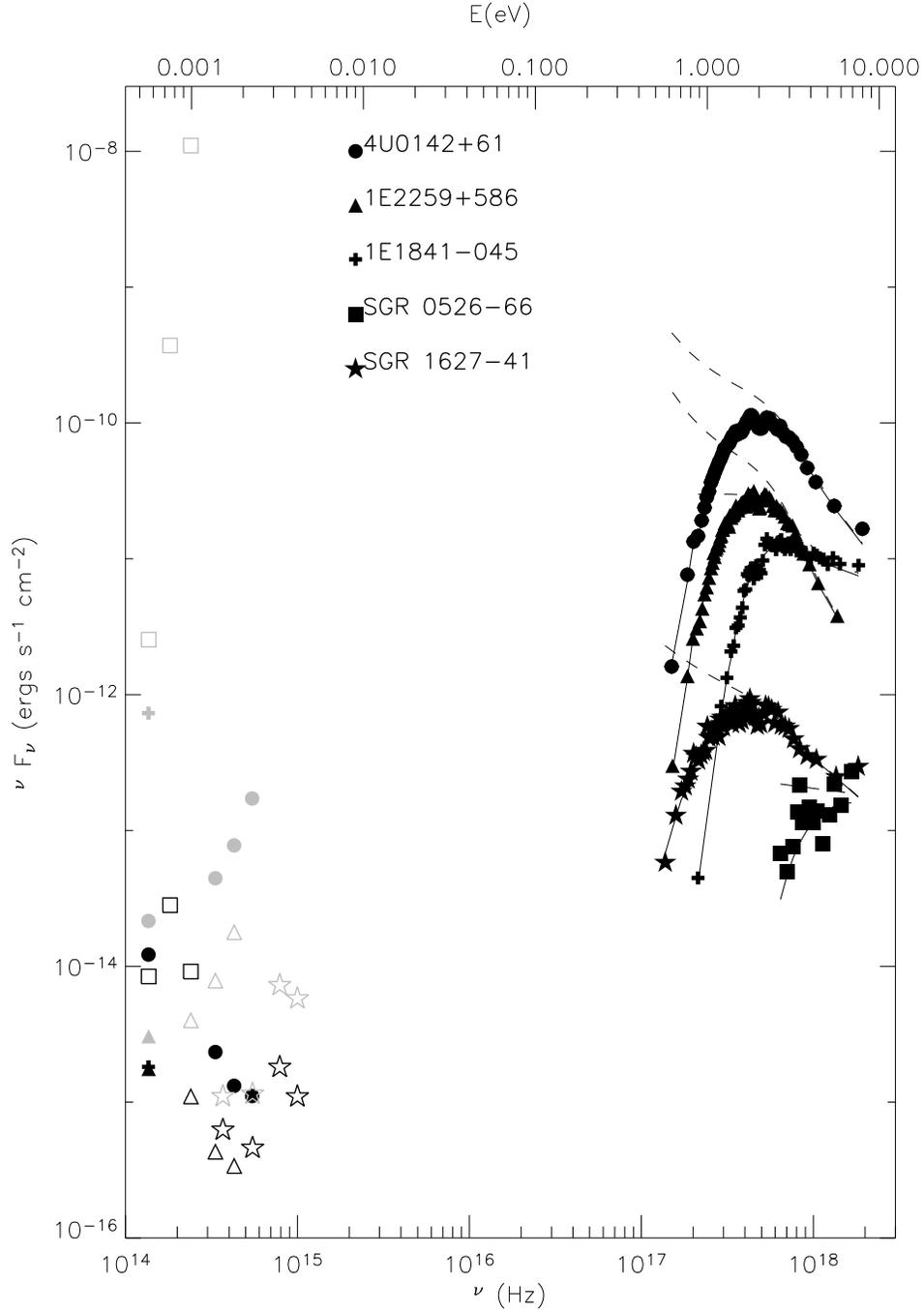}
\caption{
Broad-band spectral energy distribution covering the infra-red to the X-rays 
for several AXPs and SGRs.
Detections are identified with solid 
markers and the upper limits are denoted by the open markers. The observed  
spectra are shown in black while the unabsorbed (dereddened) spectra are indicated in 
grey.  
\label{f-spectrum} }
\end{figure}

A comparison of the broadband spectra of five AXPs and SGRs are shown
in Figure~\ref{f-spectrum}. The two SGRs shown in this figure have similar spectra
but are dimmer than the AXPs in the X-ray band, probably due to both
their larger distances (e.g., SGR~0526$-$66) and larger absorption
columns (e.g., SGR~1627$-$41). It is, therefore, expected that the
apparent optical/IR magnitudes of SGRs are significantly larger ($\sim
1.5- 2.0$) than those of AXPs if the broadband emission mechanism is
the same between the sources. We conclude that the counterparts to the
three heavily absorbed SGRs have not yet been detected and that we
would need significantly deeper observations to achieve that goal.

Such observations are useful in differentiating between different
models for AXPs and SGRs. This is because the magnitude of the
observed IR flux can place strong constraints on the nature of the
emission, even though the different variants of the magnetar model
make no concrete predictions for the IR emission. The most stringent
constraint comes from imposing the thermodynamic limit on the surface
emission from a neutron star; the maximum flux observed at infinity
from thermal processes on the surface of a neutron star is given by
the blackbody limit. For a source distance of 5~kpc and an emission 
frequency of $10^{15}$~Hz, the Rayleigh-Jeans limit to the blackbody 
approximation is
\begin{equation}
\nu F_{\nu,BB} = 5\times 10^{-18} 
\left(\frac{R_{\rm NS}}{10^6~{\rm cm}}\right)^2
\left(\frac{D}{5~{\rm kpc}}\right)^{-2}
\left(\frac{T_{\rm NS}}{0.1~{\rm keV}}\right)
\left(\frac{\nu}{10^{15}~{\rm Hz}}\right)^3
{\rm erg~s}^{-1}~{\rm cm}^{-2}, 
\end{equation}
where $R_{\rm NS}$ is the radius of the neutron star and $T_{\rm NS}$
is the temperature of the surface layer from which the IR emission
originates. In ultramagnetic surface emission models, this temperature
is typically a factor of 10 smaller than the color temperature of the
X-ray spectrum \citep{ozel01}.

For the blackbody limit to be comparable to the observed IR flux, the
temperature of the IR-emitting layers on the neutron star surface
would need to exceed 1~MeV. This temperature would not only produce an
X-ray spectrum incompatible with the data, but would also enter the regime where the
neutrino emission dominates the cooling of the atmosphere. Clearly,
the observed IR emission of the magnetars should originate in their
magnetospheres.

Studying the broadband spectra of AXPs and SGRs are crucial for
addressing the question of whether the same emission mechanism and
energy source are responsible for generating their long- and
short-wavelength emission. Observationally, the optical/IR and X-ray 
spectra cannot be fit by a simple extension of a physically plausible 
blackbody; the former rising in $\nu F_\nu$ while the latter are falling.
Moreover, the X-rays require the dissipation of magnetic energy
whereas the optical/IR luminosities can be accounted for by rotational
energy losses alone \citep{ozel04}. Clearly, a spectral break must
occur somewhere in the UV/EUV band. The position of this break and its
luminosity will provide important clues towards understanding the
energy budgets and emission mechanisms at work in AXPs and
SGRs. Searches for these sources in the UV band in conjunction with
deeper optical/IR observations are crucial in revealing the nature of
the long-wavelength emission of magnetars.

\acknowledgments

The research described in this paper was carried out at the
Jet Propulsion Laboratory, California Institute of Technology, and
at the National Space Science and Technology Center in Huntsville, and was
sponsored by NASA grant NAG5-11608. This
publication makes use of data products from the 2 Micron All Sky
Survey, which is a joint project of the University of Massachusetts
and the Infrared Processing and Analysis Center/California Institute
of Technology, funded by the National Aeronautics and Space
Administration and the National Science Foundation. It also utilized
NASA's Astrophysics Data System Abstract Service and the SIMBAD
database operated by CDS, Strasbourg, France.
F.\"O. acknowledges support by NASA through Hubble Fellowship
grant HF-01156 from the Space Telescope Science Institute, which is
operated by the Association of Universities for Research in Astronomy,
Inc., under NASA contract NAS 5-26555. P.W. and S.K.P. acknowledge support from NASA through
SAO grant GO1-2066X.

\clearpage

\clearpage

\begin{deluxetable}{clccc}
\tablecaption{Log of IR Observations
\label{t-log}}
\tablehead{\colhead{Date}& \colhead{Telescope} & \colhead{$J$ (\# images)} &
\colhead{$H$ (\# images)}& \colhead{$K_s$ (\# images)}\\ \colhead{}& \colhead{(Instr.)} &
\colhead{exptime $\times$ \#coadd}& \colhead{exptime $\times$
\#coadd}&\colhead{exptime $\times$ \#coadd} }
\startdata
\multicolumn{5}{c}{{\it \sgr}} \\ \hline
2001 March 06 & CTIO 1.5m & 9 & 9 & 9 \\ &(OSIRIS) & $60s\times 1$ & $30 s\times 1$ & $30 s\times 1$ \\
2001 March 17 & CTIO 4m   & 15 & 15 & 25 \\ &(OSIRIS)  & $30 s\times 1$ & $15 s \times 4$& $10 s \times 3$ \\
& &\nodata &\nodata & 36 \\
& &\nodata &\nodata & $10 s \times 1$ \\ \hline
& & & & \\
\multicolumn{5}{c}{{\it AXP~1E1841$-$045}} \\ \hline
2002 June 17 &ESO NTT & & & 30 \\ & (SOFI) &\nodata &\nodata &$5 s \times 12$ \\ \hline
& & & & \\
\multicolumn{5}{c}{{\it SGR~1806$-$20}} \\ \hline
2003 July 31 & ESO NTT & & & 20 \\ & (SOFI)&\nodata &\nodata &$10 s \times 6$ \\
\enddata
\end{deluxetable}

\begin{deluxetable}{cccccccc}
\tabletypesize{\small}
\tablecaption{IR Photometry in the Field of \axp \label{t-mcomps}} 
\tablehead{
\colhead{Star\tablenotemark{a}} &
\colhead{$K_s$\tablenotemark{b} } &
\colhead{$K_s$\tablenotemark{c}} &
\colhead{$K_s$\tablenotemark{d} } &
\colhead{Star} &
\colhead{$K_s$\tablenotemark{b}} &
\colhead{$K_s$\tablenotemark{c}}&
\colhead{$K_s$\tablenotemark{d}}  
}
\startdata
A & 12.89 $\pm$ 0.05& 12.96  &12.80$\pm$ 0.08 & 7 & 14.92 $\pm$ 0.08& 15.16& 14.96$\pm$0.09 \\
B & 11.99 $\pm$ 0.05& 12.08  &11.98$\pm$ 0.08& 8* & 15.36 $\pm$ 0.09& 16.36&16.05$\pm$0.25 \\
C & 14.08 $\pm$ 0.07& 14.28  &14.00$\pm$ 0.09& 9 & 15.63 $\pm$ 0.10& 15.62 &15.72$\pm$0.12 \\
D & 10.27 $\pm$ 0.05& 10.33  &10.24$\pm$ 0.08& 10& 13.87 $\pm$ 0.06& 13.94& 13.90$\pm$0.08 \\
H & 14.95 $\pm$ 0.08& 14.78  &14.91$\pm$ 0.12& 11& 11.87 $\pm$ 0.05& 11.94& 11.85$\pm$0.08\\
I & 15.54 $\pm$ 0.15& 15.18  &15.48$\pm$ 0.13& 12& 15.29 $\pm$ 0.10& 15.04& 15.32$\pm$0.11\\
L & 13.33 $\pm$ 0.06& 13.39  &13.29$\pm$ 0.09& 13& 15.20 $\pm$ 0.09& 15.27& 15.27$\pm$0.11\\
M & 15.49 $\pm$ 0.10& 15.30  &15.16$\pm$ 0.11& 14*& 15.56 $\pm$ 0.15& 13.73&15.51$\pm$0.12 \\
O & 16.01 $\pm$ 0.20& 16.30  &16.08$\pm$ 0.20& 15& 14.29 $\pm$ 0.10& 13.87& 14.29$\pm$0.09\\
P & 16.83 $\pm$ 0.25& 17.08  &16.56$\pm$ 0.25& 16*& 16.44 $\pm$ 0.15& 17.01&16.28$\pm$0.23 \\
Q & 14.37 $\pm$ 0.07& 14.41  &14.13$\pm$ 0.11& 17*& 16.11 $\pm$ 0.15& 16.78&15.83$\pm$0.13 \\
R & 15.81 $\pm$ 0.20& 15.54  &15.85$\pm$ 0.20& 18& 16.20 $\pm$ 0.20 & 16.37&16.08$\pm$0.20  \\
T & 15.69 $\pm$ 0.15& 16.04  &15.67$\pm$ 0.13& 19& 12.76 $\pm$ 0.05& 12.81 &12.73$\pm$0.08 \\
U* & 16.35 $\pm$ 0.20& 15.43 &16.07$\pm$ 0.20& 20& 16.02 $\pm$ 0.15& 16.33 &16.27$\pm$0.23 \\
V & 16.25 $\pm$ 0.20& 16.55  &\nodata        & 21*& 16.69 $\pm$ 0.20 & 15.62& \nodata  \\
Z & 15.65 $\pm$ 0.10& 15.88  &15.76$\pm$ 0.13& 22*& 15.77 $\pm$ 0.15 & 14.42&16.04$\pm$0.20 \\
1 & 16.23 $\pm$ 0.15& 16.19  &16.41$\pm$ 0.25& 23*& 16.58 $\pm$ 0.20 & 17.08&\nodata \\
2 & 12.64 $\pm$ 0.05& 12.69  &12.66$\pm$ 0.08& 24& 14.56 $\pm$ 0.07& 14.84& 14.74$\pm$0.10 \\
3 & 12.62 $\pm$ 0.05& 12.55  &12.61$\pm$ 0.10& 25& 15.75 $\pm$ 0.10& 15.69& 15.68$\pm$0.11 \\
4 & 15.40 $\pm$ 0.09& 15.24  &15.38$\pm$ 0.12& 28& 15.57 $\pm$ 0.09& 15.59& 15.34$\pm$0.10 \\
6* & 15.60 $\pm$ 0.15& 15.06 &16.00$\pm$ 0.20&   &              &      & \\
\enddata
\tablenotetext{a}{nomenclature from \citet{meregh01}}
\tablenotetext{b}{this work }
\tablenotetext{c}{from \citet{meregh01}}
\tablenotetext{d}{re-analyzed $K_s$ data set of \citet{meregh01}}
\tablenotetext{*}{Mag Diff. $> 0.5$}
\end{deluxetable}

\begin{deluxetable}{ccrrc}
\tablecaption{\chandra\ X-ray Point Sources matching 2MASS sources in
the central 6\arcmin\ around \sgr
\label{t-source6}}
\tablehead{
\colhead{RA\tablenotemark{a}} &
\colhead{DEC\tablenotemark{a}} &
\colhead{$\delta$RA (\arcsec)} &
\colhead{$\delta$DEC (\arcsec)} &
\colhead{2MASS, USNO catalog name} 
}
\startdata
16:35:29.988 & $-$47:38:05.28  & 0.768 & 0.07  & 16352991-4738053, 0375-28039446 \\ 
16:35:34.207 & $-$47:39:33.86  & 0.576 & $-$0.29     & 16353414-4739335, 0375-28043741 \\
16:35:34.666 & $-$47:33:05.86  & 0.869 & 0.47  & 16353458-4733063, 0375-28044283 \\
16:35:37.133 & $-$47:38:03.15  & 0.243 & 0.33  & 16353710-4738034, 0375-28046819 \\
16:35:37.426 & $-$47:38:49.13  & 0.930 & 0.27  & 16353733-4738494, 0375-28047088 \\
16:35:42.206 & $-$47:33:41.66  & 0.586 & 0.18  & 16354214-4733418, 0375-28052100 \\
16:35:42.341 & $-$47:34:07.00  & 0.404 & 0.26  & 16354230-4734072, 0375-28052286 \\
16:35:45.168 & $-$47:36:57.03  & 0.354 & $-$0.03    & 16354513-4736569, 0375-28055150 \\
16:35:45.494 & $-$47:37:43.93  & 0.617 & $-$0.06    & 16354543-4737438 \\
16:35:45.667 & $-$47:41:32.19  & 0.738 & $-$0.12    & 16354559-4741320, 0375-28055773: \\
16:35:47.515 & $-$47:36:08.32  & 0.384 & $-$0.03    & 16354747-4736082, 0375-28057595: \\
16:35:48.425 & $-$47:36:50.22  & 0.435 & 0.08  & 16354838-4736502, 0375-28058346: \\
16:35:49.382 & $-$47:40:08.96  & 0.728 & 0.05  & 16354930-4740090, 0375-28059500 \\
16:35:51.859 & $-$47:39:41.27  & $-$0.232 & $-$0.75  & 16355188-4739405, 0375-28062106 \\
16:35:53.477 & $-$47:32:30.68  & 0.819 & $-$0.17    & 16355339-4732305, 0375-28063751 \\
16:35:53.988 & $-$47:30:42.78  & 0.526 &  $-$0.25    & 16355393-4730425, 0375-28064268 \\
16:35:55.346 & $-$47:35:22.53  & 0.162 & 0.29  & 16355532-4735228, 0375-28065740 \\
16:35:55.764 & $-$47:39:55.75  & 1.132 & 0.24  & 16355565-4739559, 0375-28066141 \\
16:35:56.371 & $-$47:36:37.28  & 0.576 & 0.11  & 16355631-4736373, 0375-28066792 \\
16:35:58.898 & $-$47:41:02.35  & $-$1.021 &1.10& 16355899-4741034 \\
16:36:03.761 & $-$47:35:36.11  & 0.465 &  $-$0.12    & 16360371-4735359, 0375-28074553 \\
16:36:07.896 & $-$47:40:25.26  & 1.688 & 0.22  & 16360772-4740254, 0375-28078699 \\
16:36:08.969 & $-$47:32:06.17  & 1.537 &  $-$0.45    & 16360881-4732057, 0375-28079785 \\
16:36:17.681 & $-$47:32:55.41  & 0.121 & 0.27  & 16361766-4732556, 0375-28088802 \\
\enddata
\tablenotetext{a}{{\it Chandra} positions}
\end{deluxetable}

\begin{deluxetable}{lcccccc}
\tabletypesize{\small}
\tablewidth{0pt}
\tablecaption{Chandra X-ray Point Sources near SGR 1627$-$41 \label{t-allsources}}
\tablehead{
\colhead{Source} &
\colhead{RA (J2000)} &
\colhead{DEC (J2000)} &
\colhead{\chandra\ S/N\tablenotemark{a}} &
\colhead{off-axis (arcmin)} &
\colhead{Comments} \\
\colhead{} & \colhead{}& \colhead{} & \colhead{$J$} & \colhead{$H$} & \colhead{$Ks$} 
}
\startdata
1 & 16:34:43.783 & -47:35:26.42  & 6.6 & 11.2  &  no 2MASS match \\
2 & 16:34:44.542 & -47:36:47.85  & 4.5 & 11.1&  no 2MASS match\\
3 & 16:34:55.814 & -47:40:03.40  & 3.9 & 9.9 &  no 2MASS match\\ 
4 & 16:35:18.257 & -47:48:32.39  & 5.2 & 13.4&  16351836-4748321\\
  & 16:35:18.362 & -47:48:32.12  & $15.436\pm 0.057$& $14.682\pm 0.053$ & $15.196\pm$null \\
5 & 16:35:19.339 & -47:45:52.61  & 5.0 & 10.9&  no 2MASS match \\
6 & 16:35:20.534 & -47:51:09.47  & 6.0 & 15.7&  no 2MASS match \\
7 & 16:35:20.657 & -47:41:44.48  & 3.3 & 7.4 &  no 2MASS match \\
8 & 16:35:20.676 & -47:40:29.27  & 3.2 & 6.5 &  16352058-4740281\\
  & 16:35:20.589 & -47:40:28.18  &  $14.426\pm 0.079$&$13.768\pm 0.076$&$13.247\pm 0.166$ \\
9 & 16:35:20.712 & -47:28:30.69  & 9.3 &9.2& 16352057-4728317 \\
  &              &               &               &      &  AXJ1635.4-4728 ?\\ 
  & 16:35:20.580 & -47:28:31.78  & $ 13.683\pm 0.034$ &$ 12.860\pm 0.054$ & $12.453\pm 0.056$ \\
10 & 16:35:25.975 & -47:38:01.39  & 4.8 &4.4 & no 2MASS match \\
\nodata &         &               &          &                \\
\enddata
\tablenotetext{a}{0.7-7.0keV}
\tablenotetext{b}{sources with * are within the FOV of the 1.5m image}
\tablenotetext{c}{1.5m measurement}
\tablenotetext{d}{sources with ** are within the FOV of the 4m image}
\tablenotetext{e}{4m measurement}
\end{deluxetable}

\begin{deluxetable}{ccrrc}
\tablecaption{\chandra\ X-ray Point Sources matching 2MASS sources for AXP~1E1841$-$045
\label{t-sourceaxp}}
\tablehead{
\colhead{RA\tablenotemark{a}} &
\colhead{DEC\tablenotemark{a}} &
\colhead{$\delta$RA (\arcsec)} &
\colhead{$\delta$DEC (\arcsec)} &
\colhead{2MASS catalog name}
}
\startdata
18:41:09.595 & $-$4:56:32.01  & 0.015& 0.19  & 18410959-0456322 \\
18:41:26.580 & $-$4:54:25.00  & $-$0.12& $-$0.73     & 18412658-0454242 \\
18:41:27.122 & $-$4:55:24.46  & $-$0.105& 0.38  & 18412712-0455248 \\
18:41:29.348 & $-$4:58:33.81  & $-$0.435& 0.33  & 18412937-0458341 \\
18:41:30.608 & $-$4:52:25.12  & 0.300   & 0.59  & 18413058-0452257 \\
\enddata
\tablenotetext{a}{{\it Chandra} positions}
\end{deluxetable}

\begin{deluxetable}{cccccc}
\tablecaption{Photometry of sources A and B in the Field of \sgr
\label{t-phot}}
\tablehead{
\colhead{Source} &
\colhead{$J$} &
\colhead{$H$} &
\colhead{$K_s$} &
\colhead{$J-H$} &
\colhead{$H-K_s$}
}
\startdata
A & 20.4$\pm 0.5$ & 18.8$\pm 0.4$ & 18.2$\pm 0.3$ & 1.6 & 0.6\\
B & \nodata      & 19.3$\pm 0.5$ & 18.4$\pm 0.3$ & \nodata & 0.9\\ 
\enddata
\end{deluxetable}

\begin{deluxetable}{cccccc}
\tabletypesize{\scriptsize}
\tablewidth{0pt}
\tablecaption{SGR and AXP Counterparts
\label{t-axpsgrs}}
\tablehead{
\colhead{Source} &
\colhead{d (kpc)} &
\colhead{$A_V$} &
\colhead{Counterpart} &
\colhead{$K_0$} &
\colhead{ref.}
}
\startdata
\multicolumn{6}{c}{SGRs} \\ \hline
0526-66 & 50           & 1.0              & no candidate                       &            & 1 \\
1627-41 & $11.0(3)$    & 43-56            &  $K_s\gtrsim20$                    & 15--13.6   &this work, 2\\

1806$-$20 & $15.1(1.8)$  & $35(5)$, $29(2)$ & $J>21$, $H>20.5$, $K_s=18.6(1.0)$\tablenotemark{a}  & 14.6, 15.3 &3, 4\\
1900+14 & 5, 12-15     & $12.8(8)$, 19.2  & $J\gtrsim22.8$, $K_s\gtrsim20.8$   & 19.3--18.6 &5, 6\\
        &              &                  &                                    &            &  \\
\multicolumn{6}{c}{AXPs} \\ \hline
0142+614    & $>2.7$   &2.6--5.1& $V=25.6$, $R=25.0$, $I=23.8$, $K=19.6$       & 19.3--19.0 & 7, 8, 9\\
            &          &        & $J=22.3(1)$, $H=21.1(1)$, $K'=20.0(1)$       & 19.7--19.4 & 10\\ 
            &          &        & $K_s= 20.1$                                  & 19.5       & 11\\
1048.1$-$5937 & $>2.8$   & 5.8    & $J=21.7(3)$, $H=20.8(3)$, $K_s=19.4(3)$      & 18.7       & 12\\
            &          &        &  $K_s=21.1(4)$                               & 20.4       & 8 \\
1708$-$4009   & $>5$     &7.8--11.2&$H=18.85(5)$, $K'=17.53(2)$\tablenotemark{b} & 16.6--16.3 & 13 \\
            &          &        & $J=20.9(1)$, $H=18.6(1)$, $K_s=18.3(1)$\tablenotemark{b}& 17.4--17.0 & 13 \\ 
            &          &        & $H=20.43(7)$, $K'=20.00(8)$\tablenotemark{c} & 19.1--18.7 & 13 \\
1810$-$197    &  5       & 6      & $H=22.0(1)$, $K_s=20.8(1)$                   & 20.1       & 14, 15 \\ 
1841$-$045    &  6--7.5  & 14.2   & $K_s=19.4(5)$                                & 17.8       & this work, 16,17 \\
2259+586    &$\gtrsim5\pm 1$& 5.2& $K_s=21.7(2)$                               & 21.1       & 18, 19 \\
            &          &         & $K_s=20.36(15)$, $K_s=21.14(21)$            & 19.8--20.5 & 20 \\
\enddata
\tablenotetext{a}{The lack of brightening during outburst suggests that this is NOT the counterpart}
\tablenotetext{b}{Candidate A}
\tablenotetext{c}{Candidate B}
\tablerefs{(1)\citet{kap01} (2)\citet{corb99} (3)\citet{eiken01} (4)\citet{corb97} (5)\citet{vrba00}
(6)\citet{kap02b} (7)\citet{hull02} (8)\citet{dur03} (9)\citet{hull00} (10)\citet{israel03b} (11)\citet{hull04} (12)\citet{wang02} (13)\citet{israel03} (14)\citet{israel04} (15)\citet{gott04} (16)\citet{san92} (17)\citet{morii03} (18)\citet{hull01} (19)\citet{hull00a} (20)\citet{kasp03}}
\end{deluxetable}

\begin{deluxetable}{cccc}
\tablecaption{Photometry of SGR 1806$-$20 sources 
\label{t-1806}}
\tablehead{ \colhead{Source} & \colhead{$K_s$ this work} &
\colhead{$K_s$ Eikenberry} & \colhead{$K_s$ 2MASS} } 
\startdata 
A & 16.4$\pm 0.2$\tablenotemark{a} & 16.7$\pm 0.2$\tablenotemark{b} & \nodata\\
B & \nodata & 18.6$\pm 1.0$ & \nodata \\ 
C & 16.3$ \pm 0.2$ & 16.1$ \pm 0.2 $ & \nodata \\
D & 14.09$\pm 0.1$ & 13.87$\pm0.15$ & \nodata \\
E & 12.24$\pm 0.1$ & 12.00$\pm0.11$ & 12.162$\pm 0.038$ \\
\enddata \tablenotetext{a}{combined magnitude for A+B}
\tablenotetext{b}{combined magnitude for A+B is 16.53}
\end{deluxetable}

\end{document}